\documentclass[fleqn,usenatbib]{mnras}

\usepackage[T1]{fontenc}
\DeclareRobustCommand{\VAN}[3]{#2}
\let\VANthebibliography\thebibliography
\def\thebibliography{\DeclareRobustCommand{\VAN}[3]{##3}\VANthebibliography}

\usepackage{graphicx}	% Including figure files
\usepackage{amsmath}	% Advanced maths commands
\usepackage{amssymb}	% Extra maths symbols
\usepackage{appendix}
\usepackage{newtxtext,newtxmath}
\usepackage{xcolor}

\title[Low metallicity stars in TNG50]{Formation and fate of low metallicity stars in IllustrisTNG50}

\author[R. Pakmor et al.]{R\"udiger Pakmor$^{1}$\thanks{rpakmor@mpa-garching.mpg.de}, Christine M. Simpson$^{2,3}$, Freeke van~de~Voort$^4$, Lars Hernquist$^5$,  \newauthor Lieke van~Son$^{5,6,1}$, Martyna Chru{\'s}li{\'n}ska$^1$, Rebekka Bieri$^1$, Selma E.~de~Mink$^{1,6,5}$,  \newauthor Volker Springel$^{1}$\vspace*{0.15cm}\\
$^1$Max-Planck-Institut f\"{u}r Astrophysik, Karl-Schwarzschild-Str. 1, D-85748, Garching, Germany\\
$^2$Department of Astronomy \& Astrophysics, The University of Chicago, Chicago,  IL 60637, USA\\
$^3$Enrico Fermi Institute, The University of Chicago, Chicago,  IL 60637, USA\\
$^4$Cardiff Hub for for Astrophysics Research and Technology, School of Physics and Astronomy,\\ \ \ Cardiff University, Queen's Buildings, The Parade, Cardiff CF24 3AA, UK\\
$^5$Center for Astrophysics | Harvard \& Smithsonian, 60 Garden Street, Cambridge, MA 02138, USA\\
$^6$Anton Pannekoek Institute for Astronomy and GRAPPA, University of Amsterdam, NL-1090 GE Amsterdam, The Netherlands
}

\date{Accepted 2022 March 11. Received 2022 March 11; in original form 2021 December 20}

\pubyear{2021}

\begin{document}
\label{firstpage}
\pagerange{\pageref{firstpage}--\pageref{lastpage}}
\maketitle

\begin{abstract}
Low metallicity stars give rise to unique spectacular transients and are of immense interest for understanding stellar evolution. Their importance has only grown further with the recent detections of mergers of stellar mass black holes that likely originate mainly from low metallicity progenitor systems. Moreover, the formation of low metallicity stars is intricately linked to galaxy evolution, in particular to early enrichment and to later accretion and mixing of lower metallicity gas. Because low metallicity stars are difficult to observe directly, cosmological simulations are crucial for understanding their formation. Here we quantify the rates and locations of low metallicity star formation using the high-resolution TNG50 magnetohydrodynamical cosmological simulation, and we examine where low metallicity stars end up at $z=0$. We find that $20\%$ of stars with $Z_*<0.1\,\mathrm{Z_\odot}$ form after $z=2$, and that such stars are still forming in galaxies of all masses at $z=0$ today. Moreover, most low-metallicity stars at $z=0$ reside in massive galaxies. We analyse the radial distribution of low metallicity star formation, and discuss the curious case of seven galaxies in TNG50 that form stars from primordial gas even at $z=0$.
\end{abstract}


%%%%%%%%%%%%%%%%%%%%%%%%%%%%%%%%%%%%%%%%%%%%%%%%%%

%%%%%%%%%%%%%%%%% BODY OF PAPER %%%%%%%%%%%%%%%%%%

\begin{figure*}
\includegraphics[width=0.97\textwidth]{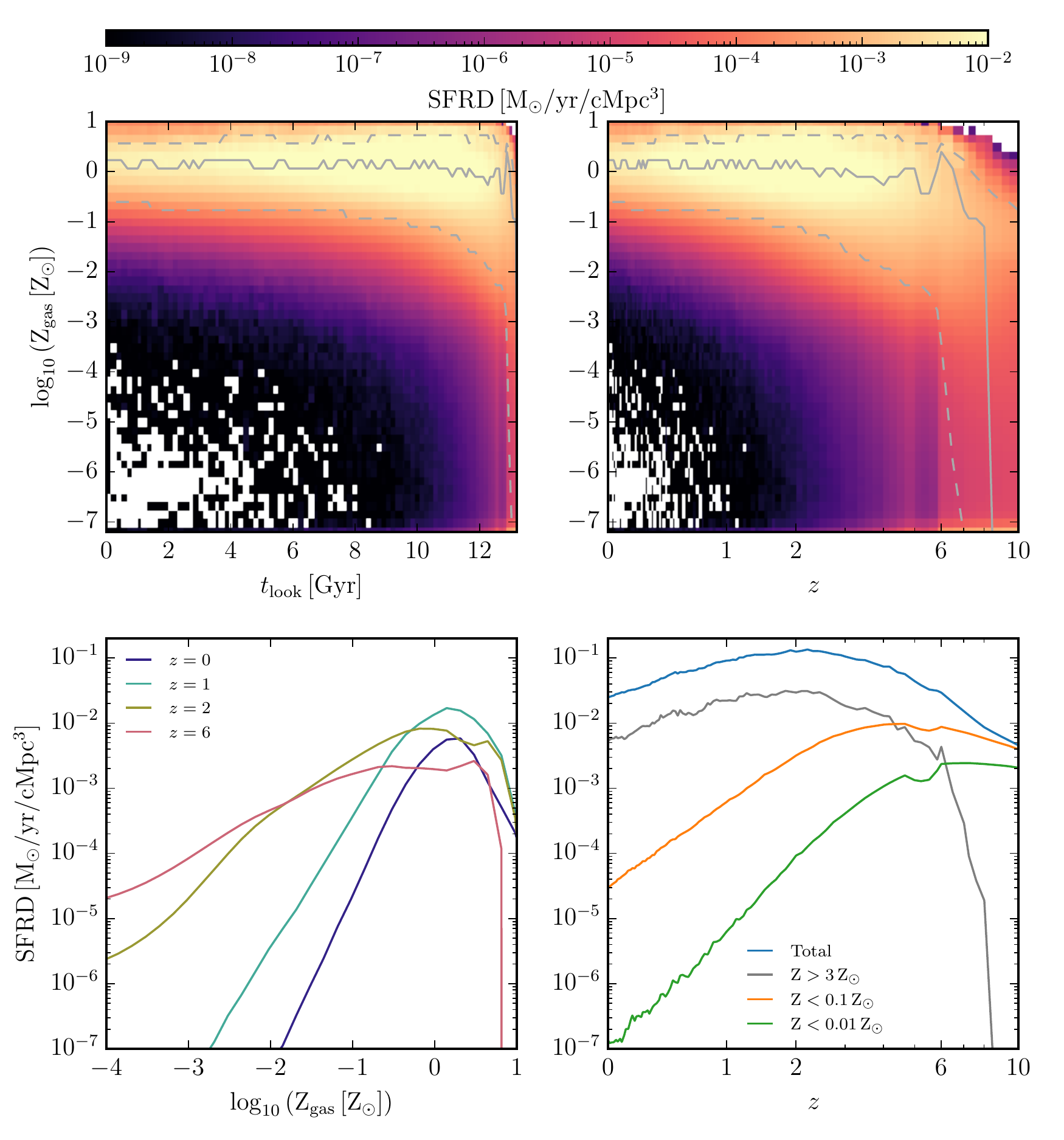}

\caption{Cosmic star formation rate density of TNG50 in bins of lookback time and gas metallicity, measured from the gas directly. The two panels in the top row show the same cosmic star formation rate density with a linear time axis (left panel) and logarithmic redshift axis (right panel). The grey solid and dashed lines show the maximum and $2\sigma$ percentiles of the distribution, respectively.
The bottom panels show slices through the same data cube. The bottom left panel shows the cosmic star formation rate density at different redshifts, the bottom right panel shows the cosmic star formation rate density integrated over metallicity for different metallicity cuts. Most stars are formed at solar metallicity after $z=6$, but stars with a metallicity as low as $10^{-2}\,\mathrm{Z_\odot}$ are still forming at $z=0$.}
\label{fig:SFR}
\end{figure*}

\section{Introduction}

Gas in the pristine Universe consists primarily of hydrogen and helium, the main products of Big Bang nucleosynthesis \citep[e.g.\ ][]{Iocco+2009,Cyburt+2016}. Over time, generations of stars enrich the interstellar medium and intergalactic space  with heavier elements, often collectively referred to as ``metals'' \citep{Burbidge+1957}. Understanding how the metallicity, that is the relative abundance of heavy elements to hydrogen, increases with time is a central question in astrophysics  \citep[e.g.\ ][]{Nomoto+2013, Maiolino+2019} that connects many different areas from stellar astrophysics to galaxy evolution and cosmology.

Stars and gas can both be studied to infer the metallicity evolution of galaxies and the Universe as a whole. Both are strongly coupled, but carry slightly different information \citep{Peeples2014,Tumlinson2017}. The stars tell us about the gas-phase metallicity at the point in time when they formed, while the gas tells us about the gas-phase metallicity at the current time.

The connection between the average gas-phase metallicity of a galaxy and its stellar mass has been studied extensively \citep[e.g.][]{Tremonti2004,KewleyEllison08,Kirby2013,Maiolino+2019} for a broad range of galaxy stellar masses. This so-called mass-metallicity relation is also present in modern numerical simulations and qualitatively consistent with observations \citep[see, e.g.][]{Lagos2016,Torrey2019,Fontanot2021}, though significant uncertainties remain between results from different observational methods \citep{KewleyEllison08,Strom2018,Cresci2019,Curti2020,Teimoorinia2021}.

In the most simplified picture the gas metallicity of a galaxy is set by cycles of self-enrichment, i.e. the stars in a galaxy produce metals and return them back into the gas phase of the galaxy, thereby enriching it and increasing the metallicity of later generations of stars. Since a significant amount of metals is produced in and released from massive stars with short lifetimes of only a few Myr, this cycle of enrichment can operate on timescales that are short compared to the evolutionary timescale of the galaxy.

Of course, in detail the processes that set the metallicity of star-forming gas in galaxies are more complicated. They include the accretion of lower metallicity gas from the circumgalactic medium surrounding galaxies; mixing between newly accreted, existing, and recently enriched gas; and galactic outflows that can carry significant amounts of metals away from the galaxy before they can be locked into new stars, but that can be reaccreted later. All of these processes, both individually and their interplay together, are an area of active research for hydrodynamical simulations \citep[e.g.][]{Vandevoort2017,Grand2019,Torrey2019,Agertz2020,Emerick2020}.

Low-metallicity stars are the focus of many studies in stellar astrophysics and cosmology.  Specifically, understanding when, where, and at what rate metal-poor stars form throughout cosmic time is critical to many active topics of research. For example, extreme stellar transients such as long gamma-ray bursts and superluminous supernovae appear to favor low metallicity environments \citep[e.g.\ ][]{Fruchter+2006, Lunnan+2014,Perley2016a,Perley2016b,Chen2017}, as well as ultra-luminous X-ray sources and the merging of massive stellar-origin black holes that are now being detected as gravitational wave sources \citep{Prestwich+2013, Abbott+2016_astrophysical_implications}.

Metallicity is very important for the evolution of stars and their death, and is thus not just a passive property, but rather a determiner of their evolutionary pathways. This is particularly true for high mass stars that lose a large fraction of their mass over a few Myrs via radiatively driven stellar winds, because the amount of mass they lose in this way decreases with decreasing metallicity \citep{Kudritzki+1987, Vink+2001, Mokiem+2007a, Vink+Sander2021}. At low metallicity, the winds are weaker and the stars retain a larger fraction of their envelope, producing more massive remnants and often more spectacular and powerful events when they die \citep[see, e.g.][]{Smartt2009,Belczynski2010,Anderson2018,Vink2018,Shen2019,Skuladottir2021}.

Moreover, metallicity also affects the interior structure of stars. In the outer layers of the stars, heavy elements are only partially ionized, making them efficient sources of opacity via bound-bound and bound-free transitions that can drive convection zones \citep[e.g.][]{Cantiello+2009} and impact the radial expansion of stars \citep[e.g.][]{Gotberg+2017}. Deeper inside stars, certain heavy elements, namely C, N and O, can act as catalysts for hydrogen fusion. The abundance of these heavy elements directly affects the efficiency of the nuclear reactions and as a result the structure of a star. Consequently, metal-poor stars are generally more compact and hotter, at least during their early evolutionary phases \citep[see, e.g.][]{Heger+2003, Gotberg+2017}.

Today star formation from low metallicity gas (i.e. significantly below solar metallicity) is significantly smaller than at higher redshift. Therefore, young low metallicity stars and the transients that follow them constitute a small fraction of all stars and transients that are visible in the local Universe. However, current surveys like E-PESSTO \citep{PESSTO} and ZTF \citep{ZTFOverview,ZTFFirstResults} and new telescopes like LSST that will become operational in the near future are discovering a large number of new transients, including more distant and very rare events from rare sources \citep{Qin2021}. Moreover, third-generation GW detectors promise to observe stellar mass Black Hole (BH) mergers out to $z >100$ \citep[e.g.][]{Sathyaprakash2019,Maggiore2020}.

To better understand the sources of events that are thought to originate from low-metallicity stars, to constrain their rates, and to find their most likely locations in the Universe, it is crucial to quantify how many low metallicity stars have formed in the Universe, how their formation rate evolves over time, and what their current formation rate is. Moreover, we want to understand which host galaxies still form low metallicity stars today, in which galaxies they have predominantly formed in the past, and where the old low metallicity stars reside today.

Since newly formed stars in the present-day Universe are dominated by solar metallicity stars, and old low metallicity stars are faint compared to young stars, observing the population of low-metallicity stars directly is challenging. Instead, we can use cosmological simulations that reproduce known constraints on the global enrichment of the Universe to learn about the properties of the population of low-metallicity stars.

There is a long history of using dark matter only simulations either in combination with analytical or semi-analytical models to constrain the low metallicity end of stars in the Milky Way and similar galaxies, often with a focus on the first generation of stars (Pop III) \citep{White2000,Hernandez2001,Diemand2005,Scannapieco2006,Brook2007,Cooper2010,Hartwig2015}.

The last several years have seen dramatic improvements to large-scale cosmological hydrodynamical simulations of galaxies. Today, state of the art simulations are able to reproduce many of the global properties of galaxies and their scaling relations for a representative portion of the universe \citep[see,e.g.][]{Vogelsberger2014Illustris,Vogelsberger2014b,Genel2014,Schaye2005Eagle,HorizonAGN2016,Dave2019Simba,TNG50Nelson}.

Such hydrodynamical models have been used already to understand in detail the evolution of the average stellar metallicity of galaxies with stellar mass, the mass-metallicity relation, gas-phase metallicity gradients in galaxies and found that the simulations are in good agreement with observations \citep[see,e.g.][]{Lagos2016,Tissera2019,Tissera2021,Torrey2019,Hemler2021}. These studies, however, focused on the average stellar and gas metallicity of galaxies. They largely ignored the low-metallicity tail, which is interesting in itself even though it is essentially irrelevant for average properties of all but the smallest galaxies.

Owing to the complex internal structure of the interstellar medium \citep[see e.g.~the recent result that indicates a significant scatter over at least one order of magnitude in the gas phase metallicity in the MW,][]{DeCia2021}, we cannot easily tell how many, if any, low-metallicity stars are still forming in typical present-day galaxies with solar metallicity by looking at their average properties. Nor can we easily constrain how many low metallicity stars a galaxy has formed in the past before it was enriched to its current metallicity.

Here, we attempt to shed light on this problem by presenting a census of low metallicity star formation in the cosmological magnetohydrodynamical simulation IllustrisTNG50 \citep[or TNG50 for short,][]{TNG50Pillepich,TNG50Nelson}. We are interested in both the instantaneous rate at which new low metallicity stars are formed, as well as the population of old low metallicity stars in galaxies at $z=0$. Besides informing us about galaxy physics, the instantaneous star formation rate in the local Universe is directly connected to transients with short delay times of up to tens of Myr, for example originating from massive stars.

We note that long delay time transients with typical delay times of Gyr are, in contrast, connected to the cumulative past formation rate of their progenitor systems. Moreover, their spatial distribution depends on the spatial distribution of old stars, which is influenced by the dynamical evolution of the host galaxy, including, for example, by galaxy mergers or internal disc instabilities.

This paper is structured as follows. We review the simulation models in Sec.~\ref{sec:simulations}. We then discuss the global metallicity dependent cosmic star formation history in Sec.~\ref{sec:global} in order to understand how many low metallicity stars the universe has formed so far, and when it formed them. In Sec.~\ref{sec:hostgalaxies}, we analyse how both the metallicity dependent star formation rate density and the total mass of low-metallicity stars at $z=0$ depend on the stellar mass of their host galaxies. We aim to understand in which galaxies low metallicity stars typically form over cosmic time, and where they reside now. We then investigate the spatial distribution of low metallicity star formation and stars in Sec.~\ref{sec:location}. Finally, we present the curious case of star formation from primordial gas at $z=0$ in Sec.~\ref{sec:zerometallicitysfr}, followed by a discussion of the implications and uncertainties of our results in Sec.~\ref{sec:discussion}, and a summary of our conclusions in Sec.~\ref{sec:summary}.

In this paper we refer to metallicity as the sum of all elements heavier than Helium, and assume a value of $Z_\odot=0.0127$ for the solar metallicity \citep{Wiersma2009, Vogelsberger2013, TNGNaiman}.

\section{Simulations}

\label{sec:simulations}

IllustrisTNG\footnote{The simulations of the IllustrisTNG project are fully publically available at \url{https://www.tng-project.org/} \citep{TNGPublicRelease}} is a set of state-of-the-art cosmological magneto-hydrodynamical simulations that evolve periodic boxes containing a representative part of the universe from high redshift to the current time \citep{TNGSpringel, TNGMarinacci, TNGNelson, TNGPillepich, TNGNaiman}. The simulations of IllustrisTNG were run with the moving-mesh code \textsc{arepo} \citep{Arepo,Pakmor2016,ArepoPublic} and evolve the equations of ideal magneto-hydrodynamics on a Voronoi-mesh coupled to self-gravity. In addition to ordinary gas, IllustrisTNG includes dark matter, stars, and black holes as collision-less particles.

IllustrisTNG employs a complex description of galaxy formation physics \citep{TNGMethodWeinberger, TNGMethodPillepich} to model the various unresolved physical processes that exchange matter, momentum and energy between gas, star particles, and black holes. The galaxy formation model is based on the Illustris \citep{Vogelsberger2014Illustris} and Auriga \citep{Grand2016} models. It includes radiative cooling of the gas in the form of primordial and metal line cooling \citep{Vogelsberger2013}, an effective subgrid model for star formation and the ISM \citep{Springel2003}, metal enrichment from stellar winds and supernovae, a heuristic model for supernova-driven galactic winds \citep{TNGMethodPillepich}, and a parameterization of the formation, growth, and feedback from supermassive black holes \citep{TNGMethodWeinberger}.

All the simulations start at $z=127$ from a homogeneous periodic box with initial density fluctuations following the \citet{Planck2016} cosmology ($\Lambda_{\Omega,0}=0.6911$, $\Lambda_{m,0}=0.3089$, $\Lambda_{b,0}=0.0486$, $\sigma_8 = 0.8159$, $n_s = 0.9667$, $h=0.6774$ ). In this work we focus on TNG50, the highest resolution box of the IllustrisTNG project \citep{TNG50Nelson,TNG50Pillepich}. The TNG50 simulation box evolves a comoving volume of $35^3\,\mathrm{Mpc/h}^3$ with a baryonic mass resolution of $5.8\times10^4\,\mathrm{M_\odot/h}$ and a dark matter mass resolution of $3.1\times10^5\,\mathrm{M_\odot/h}$ down to $z=0$. The gravitational softening length of dark matter and star particles is $200\,\mathrm{pc/h}$ at $z=0$, whereas the minimum softening of gas cells is $50\,\mathrm{pc/h}$. The combination of mass and spatial resolution reached by TNG50 is comparable to state-of-the-art zoom simulations of single galaxies \citep{Grand2016}. This high and uniform mass resolution in a large volume allows for a largely unbiased study of the full galaxy population over a large range of galaxy masses and types.

We use the Friends-of-Friends (FoF) and {\small SUBFIND} \citep{Springel2001} group finding algorithms to identify bound structures in the simulation. For this paper, we will refer to FoF groups as halos, and to gravitationally bound structures identified by {\small SUBFIND} within them as galaxies. We define the mass of a halo as $M_\mathrm{200,crit}$, i.e.~the mass contained within a sphere around the center of the halo such that the mean density in the sphere is equal to $200$ times the critical density of the universe. Moreover, we define the stellar mass of galaxies as the total mass of all star particles bound to a galaxy, and the average stellar or gas metallicity of a galaxy as the mass-weighted mean metallicity of all star particles or gas cells that are bound to the galaxy.

In TNG50, gas above a threshold density of $n_\mathrm{H}\gtrsim0.1\mathrm{cm^{-3}}$ stochastically forms stars according to the Kennicutt–Schmidt relation. Typically, when a cell forms stars the entire cell is converted to a star particle that inherits the cell's mass, momentum, and metallicity. These star particles therefore have a mass of around $8.5\times10^4\,\mathrm{M_\odot}$, the same as the baryonic mass resolution. 

Star particles are treated as average stellar populations with a Chabrier IMF \citep{ChabrierIMF} with a maximum stellar mass of $120\,\mathrm{M_\odot}$ and continuously return mass and metals to the local gas following stellar population models \citep{TNGMethodPillepich}. All stars contribute to the chemical enrichment. In practice, metal-rich gas is injected into all cells within a sphere around the star particle that contains $64$ gas cells. In other words, in TNG50,  metals newly ejected from a star particle are mixed into the $5\times10^6\,\mathrm{M_\odot}$ of gas around it. While this process takes place, the mass of the star particle is correspondingly reduced, but its metallicity is kept fixed. The maximum fraction of mass that star particles (which represent a typical stellar populations) can lose is approximately $50\%$, so no star particle is ever destroyed completely. The star particles present at $z=0$ therefore encode the complete history of star formation over the entire evolution of the universe. We note that the initial metallicity of the simulation is set to a mass fraction of $\mathrm{Z}=10^{-10}$ for all elements excluding hydrogen and helium, in order to account for unresolved early enrichment by PopIII stars, which is not explicitly followed by the simulation.

The mixing of metals is treated fully self-consistently in the framework of magnetohydrodynamics, as part of the mass fluxes between cells. The high resolution and Lagrangian nature of the TNG50 simulation reduces numerical mixing of metals due to finite resolution effects compared to lower resolution simulations or more diffusive schemes. This also allows us to better follow metallicity fluctuations in the simulation.

The IllustrisTNG simulations have originally been calibrated to reproduce the observed total cosmic star formation rate density of the Universe and the stellar mass function of galaxies \citep{TNGMethodPillepich} at the resolution of TNG100. Note that although the TNG simulations are qualitatively consistent between the three flagship runs TNG300, TNG100, and TNG50, there are systematic trends with resolution. In particular higher resolution TNG simulations systematically produce more stars in the same halos, by about $30-50\%$ at $z=0$ between TNG300 and TNG100, and galaxies at the same stellar mass are smaller in higher resolution TNG simulations \citep{TNGPillepich,Zhao2020}.

Notably relevant for this work, the TNG simulations have been shown to be consistent with the iron abundance \citep{TNGNaiman}, the sizes of galaxies and their evolution with redshift \citep{Genel2018}, and with the mass-metallicity relation of galaxies up to $z=2$ \citep{Torrey2019} for TNG100, and the metallicity gradients within galaxies at low redshift \citep{Hemler2021} for TNG50. Here we concentrate on the TNG50 simulation and discuss resolution effects specifically in Section~\ref{sec:zerometallicitysfr}.

\begin{figure}
\includegraphics[width=0.97\linewidth]{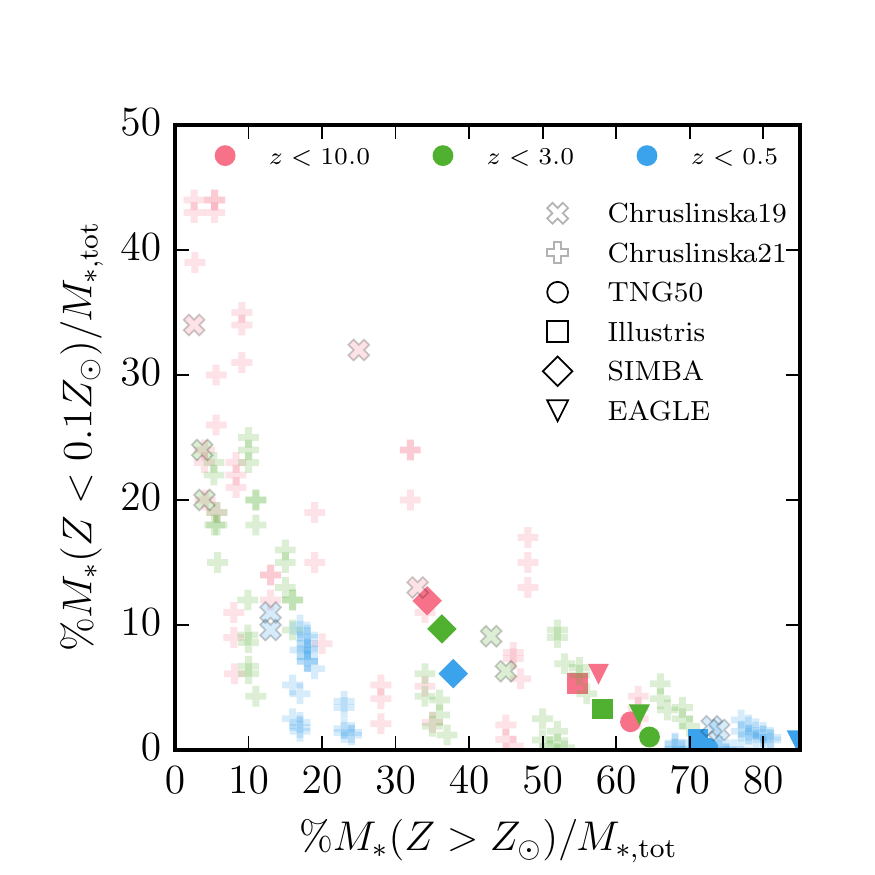}
\caption{Percentage of stellar mass with metallicity smaller than $0.1\,\mathrm{Z_\odot}$ vs. stellar mass with metallicty larger than $\mathrm{Z_\odot}$ for all stars formed at redshifts below $z=10$ (red), $z=3$ (green), and $z=0.5$ (blue). Data is shown for the most extreme observation-based models from \citet{Chruslinska2019} (crosses), the new models from \citet{Chruslinksa2021} (plusses) and for the simulations TNG50 (circles), Illustris \citep{Vogelsberger2014Illustris} (squares), Simba \citep{Dave2019Simba} (diamonds), and Eagle \citep{Schaye2005Eagle} (triangles). The various cosmological simulations are generally consistent with each other and with some of the observation-based models.}
\label{fig:StellarMassMetallicity}
\end{figure}

\begin{figure*}
\includegraphics[width=0.97\textwidth]{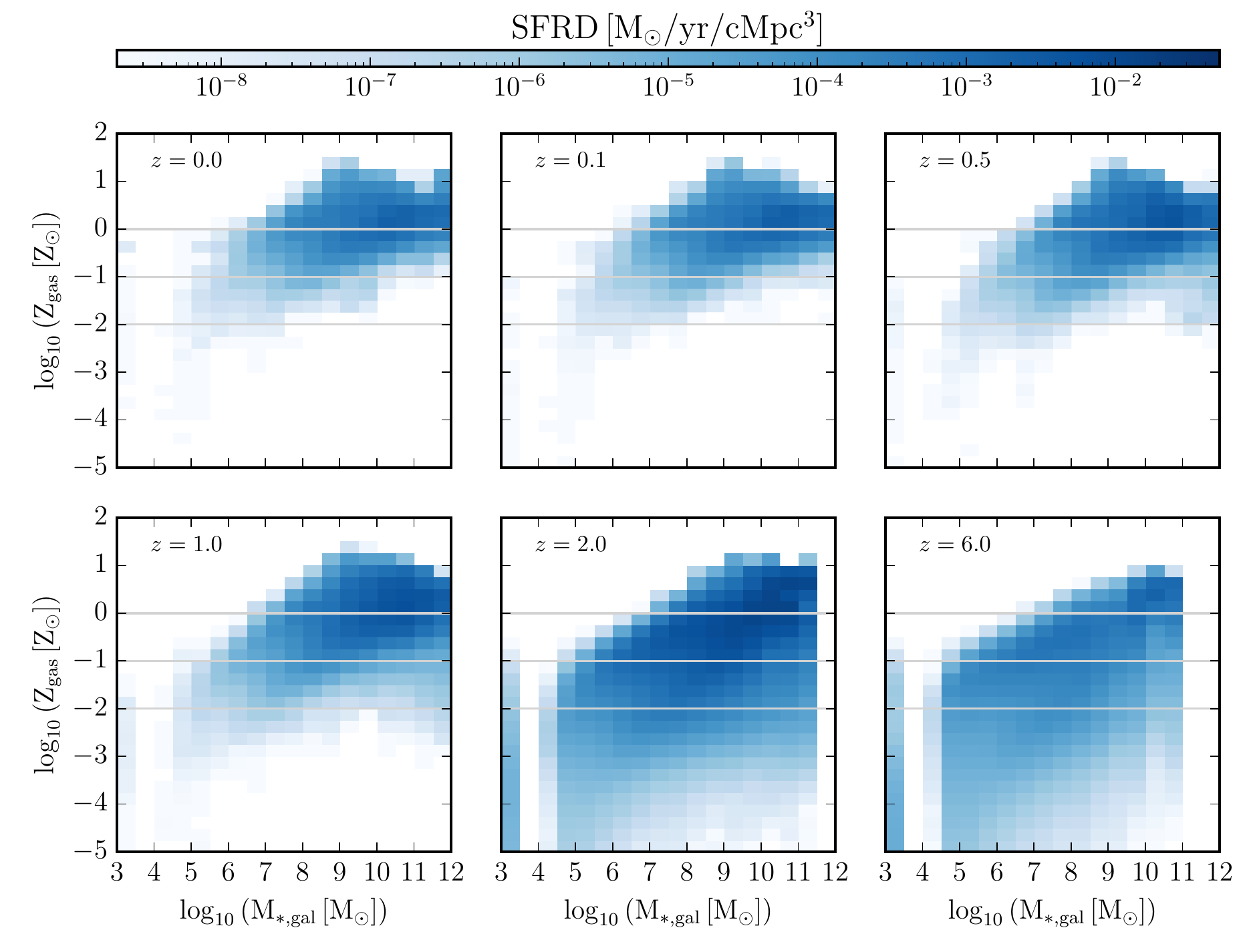}
\caption{Cosmic star formation rate density at different redshifts in bins of the metallicity of the star-forming gas and the stellar mass of the host galaxy. Galaxies without stars are set to a stellar mass of $M_\mathrm{*,gal}=10^3\,\mathrm{M_\odot}$. Horizontal grey lines indicate metallicities of $Z_\odot$, $0.1\,Z_\odot$, and $0.01\,Z_\odot$. At $z>2$ stars of arbitrarily low metallicity form in galaxies of any mass. After $z=1$ the formation of low metallicity stars becomes gradually restricted to low mass galaxies.}

\label{fig:SFRHostMass}
\end{figure*}

\section{How does the cosmic star formation history depend on metallicity?}

\label{sec:global}

We first examine the evolution of the global star formation rate with time, split up by metallicity, in order to quantify how many low metallicity stars are formed in TNG50 overall. The low metallicity star formation rate at any given time is directly connected to the rate of short delay time transients from low metallicity channels. Moreover, the cumulative number of stars formed at low metallicity up to a certain redshift will set the rate of long delay time transients from low metallicity channels at this redshift.

The evolution of the global star formation rate density with time and metallicity in the full TNG50 simulation box is shown in the top panels of Fig.~\ref{fig:SFR}. We display it twice, with a linear time axis (top left panel), and a logarithmic redshift axis (top right panel) to emphasize different epochs of the evolution of the simulated universe. The star formation rate density is computed as the sum of the instantaneous star formation rates of all cells at a given  time, i.e.~each vertical column is computed from one time slice of the simulation. Note that $1.8\times10^{-3}$ of the star formation at redshifts $z<2$ happens at a metallicity $Z>10\,\mathrm{Z_\odot}$.

The very first star particles form with the initial metallicity in the centers of the first galaxies. The first massive stars then quickly enrich their surroundings, and later generations of stars are born at higher and higher metallicity. This self-enrichment proceeds rapidly in the early universe. Shortly after reionisation at $z=6$ the peak of the metallicity distribution of newly formed stars is already at a metallicity $Z>0.1\,\mathrm{Z_\odot}$. It reaches solar metallicity around a redshift of $z=2$ and remains unchanged until $z=0$, consistent with semi-numerical enrichment models \citep{Ucci2021}.

There is still a significant contribution of low metallicity gas to the total star formation rate that slowly decreases with time. Even at $z=0$, a non-zero amount of star formation occurs at metallicities smaller than $0.01\,\mathrm{Z_\odot}$. Note that the small decrement in the star formation rate density after $z=6$ at low metallicities $Z<10^{-3}\,\mathrm{Z_\odot}$ is caused by reionisation heating up gas in low mass galaxies and suppressing their star formation until they grow more massive and self-shielding becomes effective.

As  explicitly shown in the lower left panel of Fig.~\ref{fig:SFR}, the peak of the metallicity dependent star formation rate density remains close to solar metallicity from $z=2$ to $z=0$. The distribution is not symmetric around its peak at solar metallicity, but has a much longer tail to lower metallicities than to higher metallicities. The tail to low metallicities is well described by a power-law that becomes significantly steeper at later times.

The time dependent cosmic star formation rate density for all gas, as well as for several metallicity cuts, is shown in the lower right panel of Fig.~\ref{fig:SFR}. The star formation rates for all these sets increase with time at high redshift, peak at different times, and then decrease towards $z=0$. The peak is reached earlier for lower metallicities. In particular, star formation at low metallicity ($\mathrm{Z_{gas}}<0.1\,\mathrm{Z_\odot}$) peaks around redshift $z=4$, and star formation at very low metallicity ($\mathrm{Z_{gas}}<0.01\,\mathrm{Z_\odot}$) already peaks at $z=6$, close to reionisation. Note that although the star formation rate at $\mathrm{Z_{gas}}<0.1\,\mathrm{Z_\odot}$ is about two orders of magnitude lower at $z=0$ compared to its peak around $z=4$, about $20\%$ of the stars with $\mathrm{Z_{*}}<0.1\,\mathrm{Z_\odot}$ in TNG50 are formed after $z=2$, compared to $40\%$ between $z=4$ and $z=2$, and another $40\%$ before $z=4$. Very low metallicity stars, in contrast, form predominantly at high redshift, and only $3\%$ of all stars with $\mathrm{Z_{*}}<0.01\,\mathrm{Z_\odot}$ form after $z=2$ in TNG50. All fractions were computed from the initial stellar mass formed, but change by less than $1\%$ when we compute them for the stellar masses of stars that survived until $z=0$.

At $z=0$ a fraction of about $10^{-3}$ of the ongoing star formation in TNG50 happens in gas with a metallicity $\mathrm{Z}<0.1\,\mathrm{Z_\odot}$, and only $10^{-5}$ of it has a metallicity of $\mathrm{Z}<0.01\,\mathrm{Z_\odot}$. In contrast, about $20\%$ of the ongoing star formation happens with a metallicity of $\mathrm{Z}>3\,\mathrm{Z_\odot}$ in TNG50 at $z=0$.

We compare the amount of low and high metallicity stars formed in TNG50 to other state-of-the-art cosmological galaxy simulations \citep{Vogelsberger2014Illustris,Schaye2005Eagle,Dave2019Simba} and observation-based models using empirical scaling relations \citep{Chruslinska2019,Chruslinksa2021} in Fig.~\ref{fig:StellarMassMetallicity}.  We can see that all cosmological simulations are qualitatively consistent with each other. Among the considered simulations, particularly Illustris, TNG50, and EAGLE agree well with each other. Notably EAGLE produces a significantly larger fraction of its stars with larger than solar metallicity after a redshift of $z=0.5$. SIMBA forms stars at systematically lower metallicities than the other simulations. 

The cosmological simulations have a much smaller mass fraction of low metallicity $\mathrm{Z}<0.1\,\mathrm{Z_\odot}$ stars, and a higher fraction of stars with a metallicity larger than $\mathrm{Z_\odot}$ than the analytical models of \citet{Chruslinska2019}. In contrast the more recent models discussed in \citet{Chruslinksa2021} are broadly consistent with cosmological simulations. In these models \citet{Chruslinksa2021} use the redshift-invariant fundamental metallicity relation \citep[e.g.][]{Ellison08,Mannucci10} to characterise the metallicity of galaxies, which leads to a much more gradual metallicity evolution at redshift $\gtrsim$2 than what is inferred from the extrapolated evolution of the mass-metallicity relation used in \citet{Chruslinska2019}. Such slower metallicity evolution is more in line with theoretical predictions. 

In the simulations, however, efficient self-enrichment of star-forming gas apparently prevents a large number of low metallicity stars from being formed, and any efficiently star-forming galaxy quickly reaches (super-)solar metallicity in its center. Assuming a solar metallicity of $\mathrm{Z_\odot}=0.02$ instead of the value used in TNG internally of $\mathrm{Z_\odot}=0.0127$ does not significantly change these results. Notably the percentage of star formation at super-solar metallicities drops by $\sim20\%$ for all hydro sims when switching to $\mathrm{Z_\odot}=0.02$. A more detailed comparison between cosmological simulations and observation-based models would be worthwhile to learn from their differences, but is beyond the scope of this paper.

%%%%%%%%%%%%%%%%%%%%%%%%%%%%%%%%%%%%%%%%%%%%%%%%%%%%%%%%%%%%%%%%%%%%%%
\section{In which host galaxies do metal poor stars form and where are the old low-metallicity stars?}
%%%%%%%%%%%%%%%%%%%%%%%%%%%%%%%%%%%%%%%%%%%%%%%%%%%%%%%%%%%%%%%%%%%%%%

\label{sec:hostgalaxies}

\begin{figure}
\includegraphics[width=0.97\linewidth]{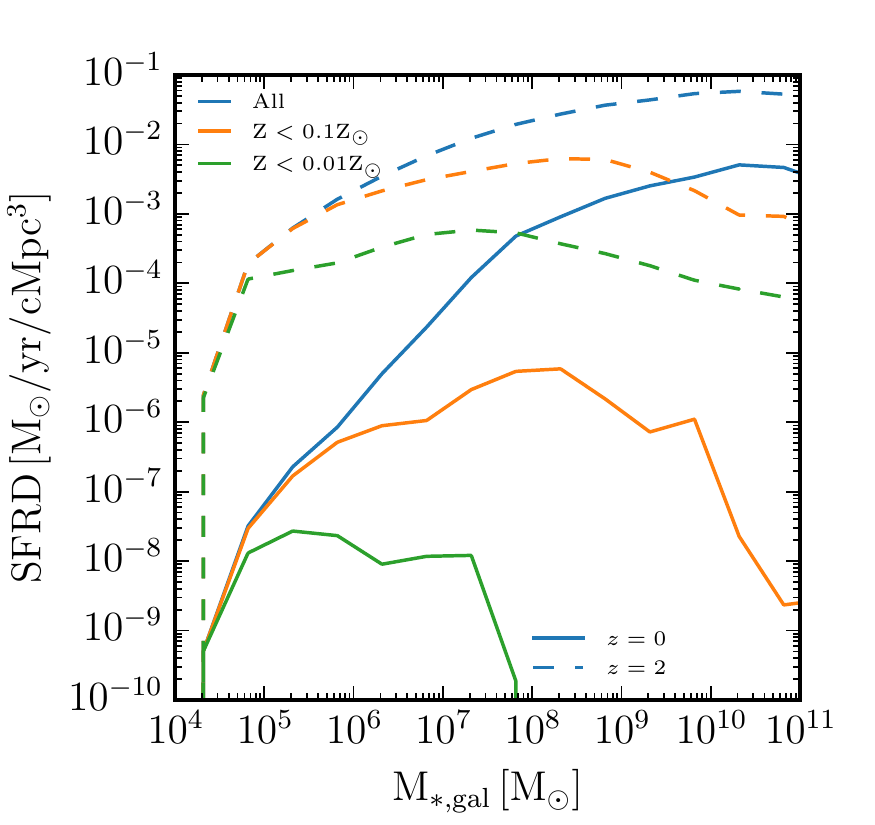}
\caption{Cosmic star formation rate density at redshifts $z=0$ and $z=2$} versus galaxy stellar mass for different metallicity cuts. For $z=2$ the star formation rate density is essentially flat in galaxy mass. At $z=0$ very low metallicity star formation is limited to low mass galaxies.
\label{fig:SFRHostMassSlices}
\end{figure}

\begin{figure}
\includegraphics[width=0.97\linewidth]{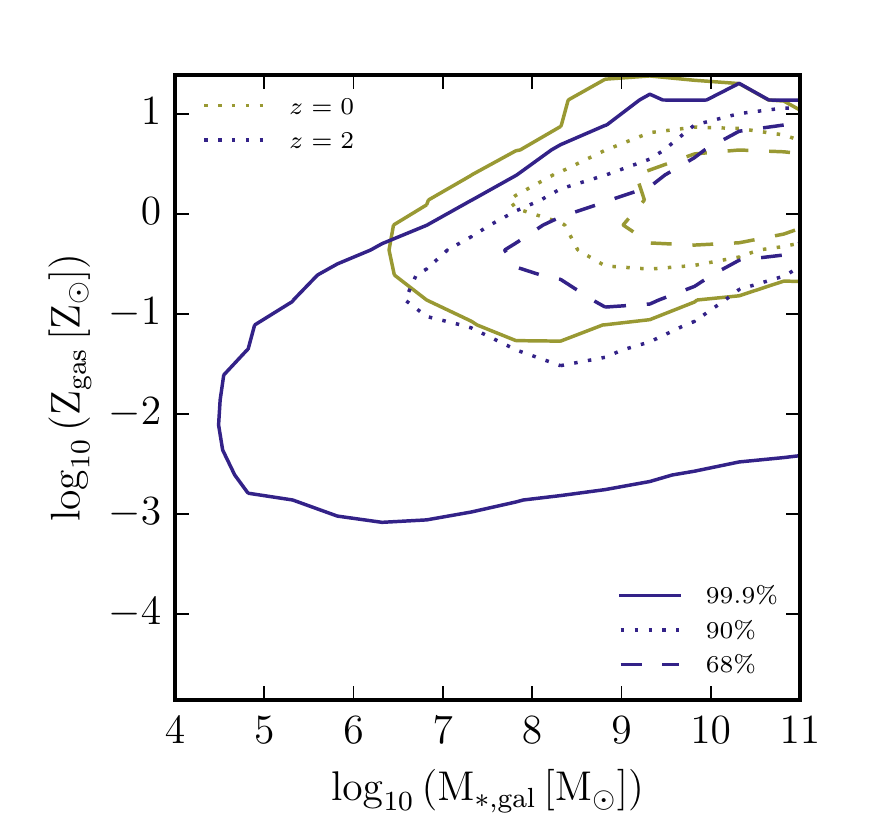}
\caption{Contours of $68\%$, $90\%$, and $99.9\%$ of the cosmic star formation rate density at redshifts $z=0$ and $z=2$ versus galaxy stellar mass. The star formation rate density contours are dominated by solar metallicity stars already at high redshift and shrink towards solar metallicity star formation in massive galaxies at $z=0$.}
\label{fig:SFRHostMassContours}
\end{figure}

Going beyond the global star formation history we wish to understand in which host galaxies low metallicity stars form and in which galaxies they reside now at $z=0$. We first look at the host galaxies of ongoing star formation at any redshift in Sec.~\ref{sec:hostgalaxiessfr}. These galaxies potentially host short delay time transients. We then look at the host galaxies of low metallicity stars of any age at $z=0$ in Sec.~\ref{sec:hostgalaxiesstars} to get an idea of where we expect to find long delay time transients.

\begin{figure}
\includegraphics[width=0.97\linewidth]{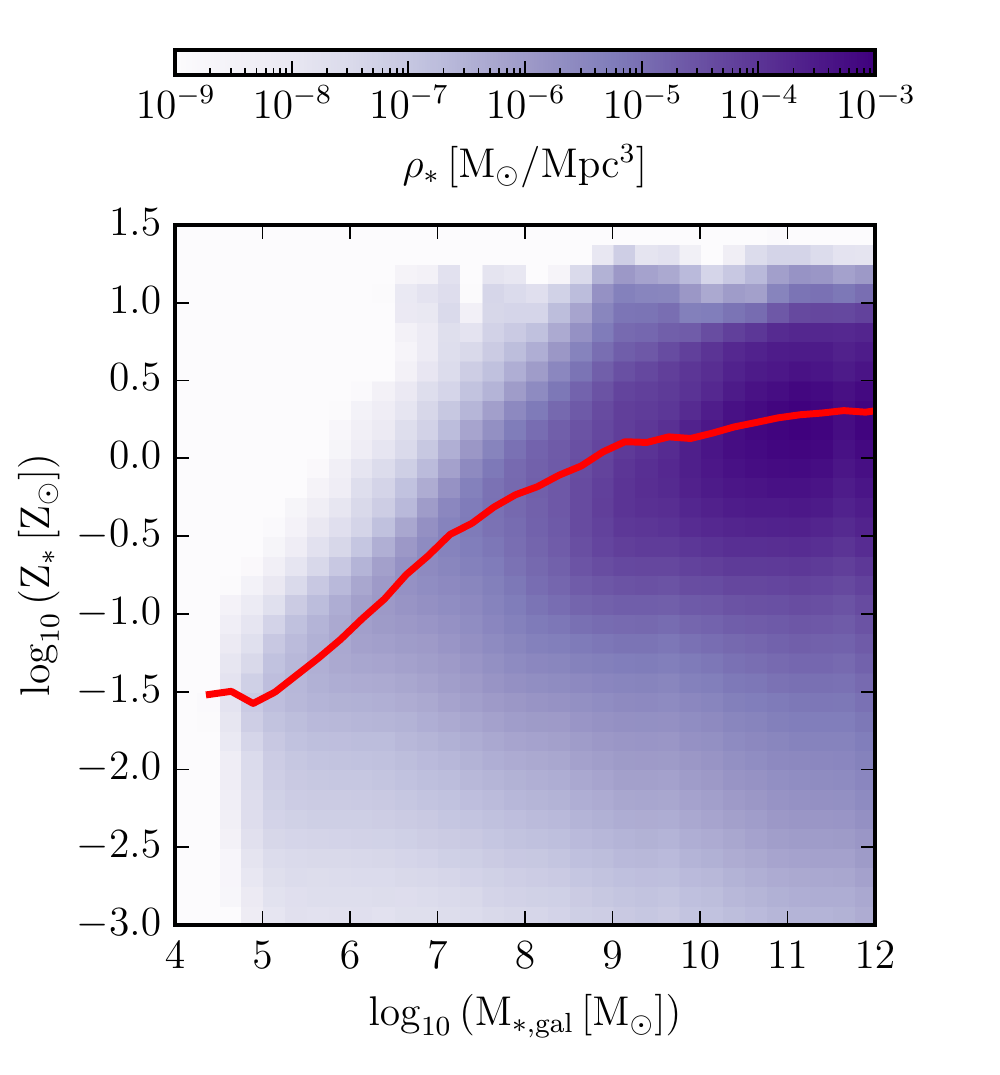}
\caption{Stellar mass density in bins of stellar metallicity and total stellar mass of the host galaxy at $z=0$. The red line shows the average stellar metallicity of all stars in all galaxies of a given stellar mass. Most low metallicity stars live in massive galaxies at $z=0$.}
\label{fig:StellarMassInGalaxies}
\end{figure}

\begin{figure}
\includegraphics[width=0.97\linewidth]{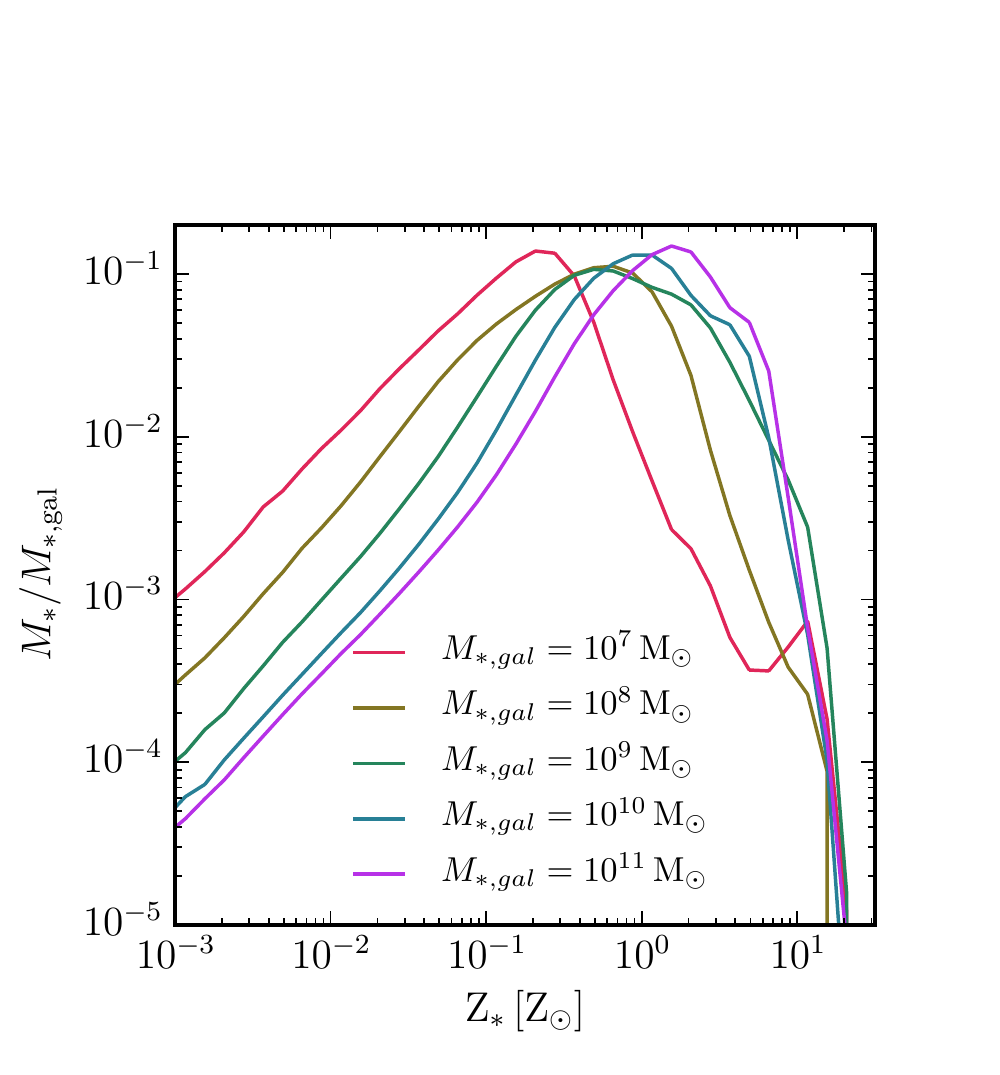}
\caption{Average relative contribution of stars at different metallicity to the total stellar mass of galaxies at different masses at $z=0$. The lines shown are identical to normalised vertical slices through Fig.~\ref{fig:StellarMassInGalaxies}. The peak of the metallicity distribution of stars moves to lower metallicity with smaller galaxy mass. All galaxies have a similar low metallicity tail.}
\label{fig:StellarMassInGalaxiesSlices}
\end{figure}

\begin{figure}
\includegraphics[width=0.97\linewidth]{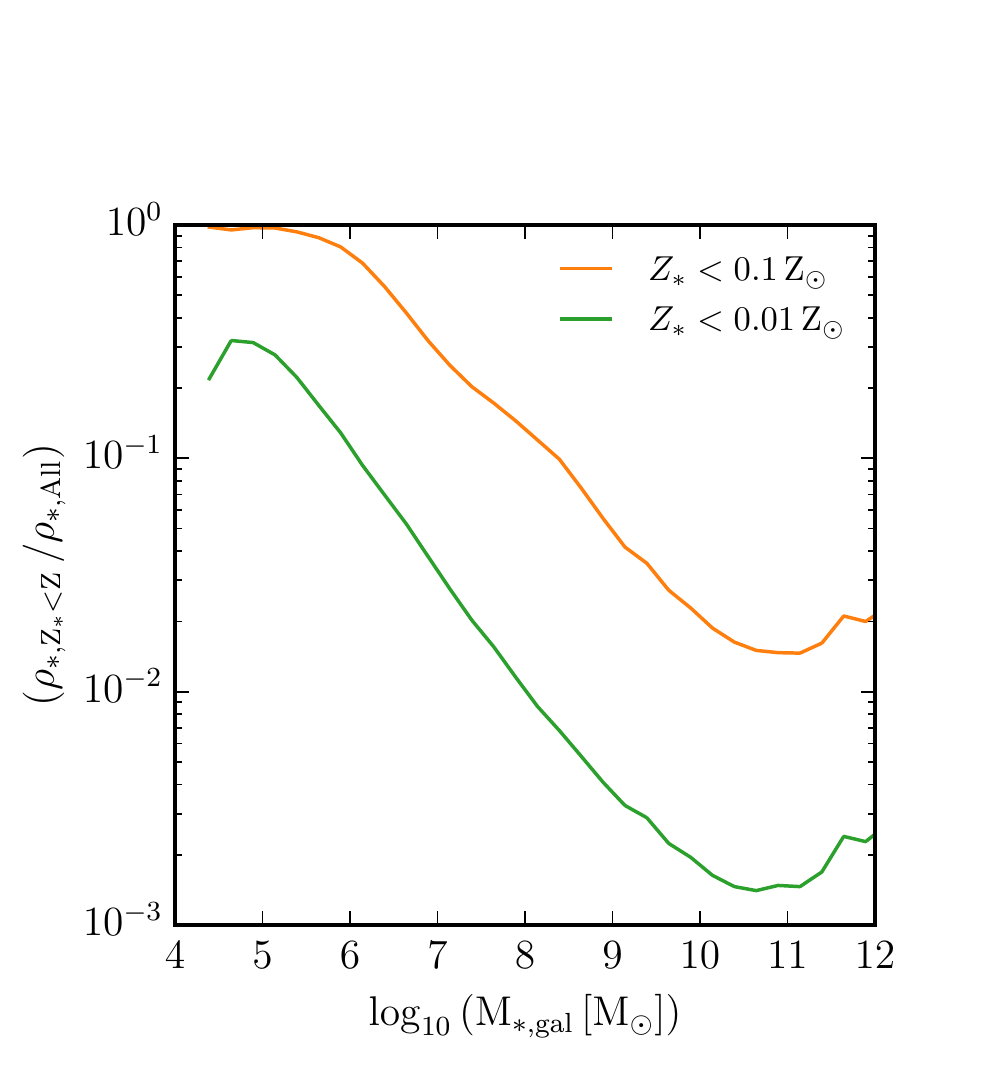}
\caption{Fraction of stellar mass density below different metallicity thresholds relative to the total stellar mass density, as a function of the total stellar mass of host galaxies at $z=0$. Low metallicity stars contribute negligible mass to the total stellar mass of massive galaxies at $z=0$ but dominate low mass galaxies.}
\label{fig:StellarMassInGalaxiesSum}
\end{figure}

\subsection{The host galaxies of metal poor star-formation}

\label{sec:hostgalaxiessfr}

We first examine how ongoing star formation at various redshifts is distributed over host galaxies, as characterised by their stellar mass and gas metallicities shown in Fig.~\ref{fig:SFRHostMass}. Here the stellar mass serves as a proxy for the primary properties of a galaxy. For orientation in the plot, let us remind ourselves that the Milky Way has a stellar mass of about $5\times10^{10}\mathrm{M_\odot}$ \citep{Cautun2020}.

The distribution of star formation over galaxies with different stellar mass allows us to better understand which galaxies form low metallicity stars at what times. It also allows us to determine where short delay time transients from low metallicity stars can be expected. We aim to understand which galaxies host any low metallicity star formation, which galaxies dominate the overall production of low metallicity stars, and how this changes with time.

At $z=2$ and larger redshifts (middle and right panels of the bottom row of Fig.~\ref{fig:SFRHostMass}), the parameter space of star formation versus gas metallicity and stellar mass of the host galaxy is almost fully populated. The exceptions are galaxies with small stellar mass and high metallicity that do not exist in TNG50. The smaller the stellar mass of a galaxy, the smaller is the upper limit on the metallicity it can have and this limit does not change significantly with time for galaxies with $M_\mathrm{*,gal}<10^9\,\mathrm{M_\odot}$. This is consistent with the assumption that self-enrichment is the critical ingredient to increase the average metallicity of galaxies.

The lowest mass galaxies with a stellar mass of $M_\mathrm{*,gal}<10^6\,\mathrm{M_\odot}$ only form stars with a metallicity up to $0.1\,\mathrm{Z_\odot}$. These galaxies did not produce many stars yet. Moreover, low-mass galaxies can lose a large fraction of metals via galactic winds because their gravitational potential wells are shallow. Note, however, that these galaxies are barely resolved with fewer than $100$ star particles and their properties should be interpreted with caution.

Slightly more massive galaxies with $M_\mathrm{*,gal}=10^7\,\mathrm{M_\odot}$ can form stars with a metallicity up to $\mathrm{Z_\odot}$, and galaxies with stellar masses $M_\mathrm{*,gal}>10^9\,\mathrm{M_\odot}$ already form the most enriched stars that will ever be formed by $z=2$. For these massive galaxies most stars form around $\mathrm{Z_\odot}$ already then, but the metallicity of their star-forming gas covers a wide range, going from a metallicity as high as $10\,\mathrm{Z_\odot}$ to a metallicity as low as $10^{-4}\,\mathrm{Z_\odot}$ at $z=2$. The tail to very low metallicity star formation even in massive galaxies at this time may be the result of accretion of very low metallicity gas directly into the star-forming phase of those galaxies. 

At $z=1$, star formation at metallicities $\mathrm{Z}<10^{-3}\,\mathrm{Z_\odot}$ has essentially ceased except for the smallest star-forming galaxies that only now begin to form stars for the first time. At this time, massive galaxies have enriched their environment substantially, and low metallicity gas that is accreted from their environment is mixed with enriched gas before it can form stars.  This establishes a floor to the metallicity of star-forming gas. 

This lower limit on the metallicity of newly formed stars moves up to $10^{-2}\,\mathrm{Z_\odot}$ at $z=0.5$. Interestingly, the metallicity floor is mostly independent of the galaxy stellar mass at $z=1$ and $z=0.5$. But at $z=0$ the metallicity floor of star-forming gas shows a clear dependence on galaxy stellar mass. The highest mass galaxies with stellar mass $M_\mathrm{*,gal}=10^{10}\,\mathrm{M_\odot}$ have a metallicity floor of $0.1\,\mathrm{Z_\odot}$, whereas the smallest galaxies can still form stars with a metallicity as low as $10^{-2}\,\mathrm{Z_\odot}$.

Note that at all times there exist galaxies that are star-forming but have not formed any stars yet.
This difference arises from the stochastic nature of our star formation algorithm. In other words the time-integrated star formation rate of these galaxies is smaller or comparable to the mass of a single star particle. While the existence of galaxies that can form stars but have not done so yet is plausible at high redshift, there is no observational evidence for galaxies that have stars today but did not form any stars before reionisation at $z=6$. We discuss the numerical and model limitations that potentially affect these galaxies in detail in Sec.~\ref{sec:zerometallicitysfr}.

To get a better quantitative understanding of which galaxies dominate low and high metallicity star formation at different redshifts, we show the total star formation rate density for a given stellar mass of the host galaxies and a metallicity below $0.01\,\mathrm{Z_\odot}$, below $0.1\,\mathrm{Z_\odot}$, or without a metallicity cut for different redshifts in Fig.~\ref{fig:SFRHostMassSlices}. We see that at $z=2$ and before, low metallicity stars with $Z<0.1\,\mathrm{Z_\odot}$ are formed in approximately equal amounts in all galaxies, independent of their stellar mass. Note, however, that massive galaxies form many more high metallicity stars than low mass galaxies, so the fraction of stars formed at low metallicity is much smaller for massive galaxies than for low mass galaxies. At $z=1$ this still holds for most galaxy masses, even though the difference in total star formation rate has become much larger between massive and low mass galaxies. 

At $z=0$, star formation at $\mathrm{Z}<0.1\,\mathrm{Z_\odot}$ is dominated by low-mass galaxies with $M_\mathrm{*,gal}\lesssim 10^8\,\mathrm{M_\odot}$. Star formation at very low metallicity of $\mathrm{Z}<10^{-2}\,\mathrm{Z_\odot}$ is limited to galaxies with a stellar mass below $M_\mathrm{*,gal}\lesssim 10^{7.5}\,\mathrm{M_\odot}$, but is in this range still roughly independent of galaxy mass. On the other end of the metallicity distribution, stars with very high metallicity $\mathrm{Z}>10\,\mathrm{Z_\odot}$ are predominantly formed in galaxies with $M_\mathrm{*,gal}\gtrsim 10^9\,\mathrm{M_\odot}$, as shown in Fig.~\ref{fig:SFRHostMass}.

To better quantify the contribution of low metallicity star formation to the total amount of star formation we show contours enclosing $68\%$, $90\%$, and $99.9\%$ of the star formation rate density at different redshifts in Fig.~\ref{fig:SFRHostMassContours}. The peak of the star formation rate density distribution moves remarkably little from $z=2$ to $z=0$ for massive galaxies with $M_\mathrm{*,gal}\sim 10^{10}\,\mathrm{M_\odot}$. For these galaxies it only becomes more concentrated around solar metallicity, which can also be seen for all redshifts in Fig.~\ref{fig:SFRHostMass}. At $z=2$, the $90\%$ contour still includes gas with a metallicity below $0.1\,\mathrm{Z_\odot}$, while at $z=0$ the lowest metallicity included in the $90\%$ contour is about $\log{\mathrm{Z/Z_\odot}}=-0.5$. This again emphasizes the dominance of star formation at about solar metallicity. Note that as a result the global scaling relations of galaxies, like for example the mass-metallicity relation, will essentially be insensitive to the total amount of low metallicity stars formed. Note also that the $99.9\%$ contour at $z=2$ extends down to a metallicity of $10^{-3}\,\mathrm{Z_\odot}$ and a galaxy mass of $M_\mathrm{*,gal}=10^{5}\,\mathrm{M_\odot}$. It shrinks substantially until $z=1$ to about the size the $90\%$ contour had at $z=2$, and slightly further until $z=0$ when it does not include any star formation below a metallicity of $0.1\,\mathrm{Z_\odot}$ anymore.

\begin{figure*}
\includegraphics[width=0.97\textwidth]{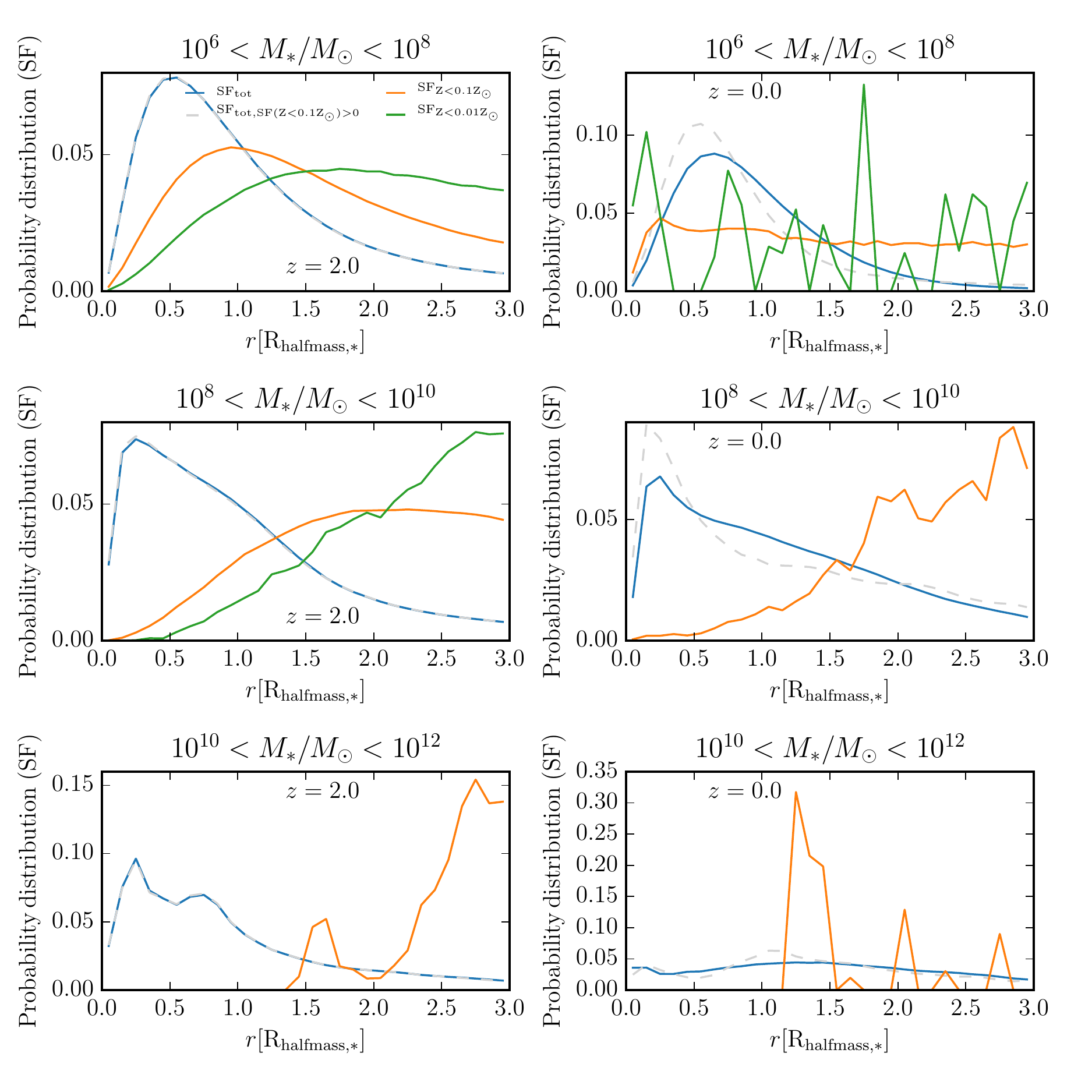}

\caption{Stacked normalised radial profiles of star formation at $z=2$ (left column) and $z=0$ (right column) for all galaxies in different stellar mass ranges (rows). The profiles are first normalised to the stellar half mass radius of the individual galaxies, then stacked. The dashed gray lines also show the profile of all star formation, but limited to the subset of galaxies that have non-zero ongoing star formation with $Z<0.1\,\mathrm{Z_\odot}$. Low metallicity stars are biased to form at larger radii than average stars.}
\label{fig:profilesSF}
\end{figure*}

\begin{figure}
\includegraphics[width=0.97\linewidth]{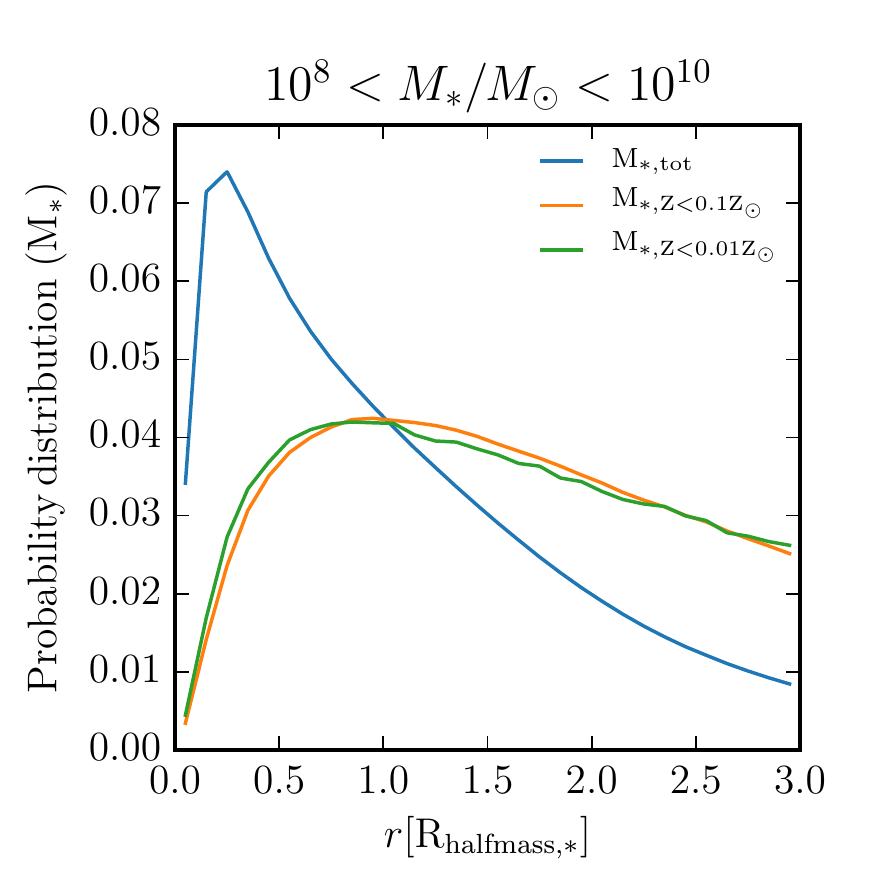}

\caption{Stacked normalised radial profiles of stellar mass at $z=0$ for galaxies with a stellar mass between $10^8\,\mathrm{M_\odot}$ and $10^{10}\,\mathrm{M_\odot}$. The profiles are first normalised to the stellar half mass radius of the individual galaxies, then stacked. Low metallicity stars are biased towards larger radii at $z=0$.}
\label{fig:profilesMass}
\end{figure}

%%%%%%%%%%%%%%%%%%%%%%%%%%%%%%%%%%%%%%%%%%%%%%%%%%%%%%%%%%%%%%%%%%%%%%
\subsection{The host galaxies of metal poor stars at $z=0$}
%%%%%%%%%%%%%%%%%%%%%%%%%%%%%%%%%%%%%%%%%%%%%%%%%%%%%%%%%%%%%%%%%%%%%%

\label{sec:hostgalaxiesstars}

After looking at the host galaxies of ongoing metal poor star formation, we now turn to the host galaxies of metal poor stars of any age at $z=0$. As seen above, those are dominated in number by old stars. Once we have an idea of how many low metallicity stars a galaxy with a given stellar mass typically contains today, we can also turn the question around and estimate how likely it is that a transient that we observe in a galaxy with known stellar mass originated from a low metallicity progenitor system.

To this end we show the distribution of stars at different metallicities among galaxies of different stellar masses at $z=0$ in Fig.~\ref{fig:StellarMassInGalaxies}. Note that this distribution not only depends on the full history of star formation of each individual galaxy, but also on its merger history. The results show not only that most stars are in the most massive galaxies in TNG50, but also that this is true for stars at any metallicity. In other words, stars at any given metallicity are preferentially part of more massive galaxies, rather than small galaxies, even for low metallicities, consistent with previous results \citep{Artale2019,Artale2020,Chruslinksa2021} and despite the large number of small galaxies that have a low average metallicity. 

For galaxies with a stellar mass between $10^7\,\mathrm{M_\odot}$ and $10^{11}\,\mathrm{M_\odot}$, we show the average relative contribution of stars at different metallicities to the total stellar mass at $z=0$ in five mass bins in Fig.~\ref{fig:StellarMassInGalaxiesSlices}. As expected from the mass-metallicity relation for galaxies, the peak shifts slightly to lower metallicity with lower stellar mass of the galaxy. While the slope towards low metallicity is very similar for all galaxy masses shown here, the normalisation changes slightly, i.e. galaxies with a stellar mass of $10^9\,\mathrm{M_\odot}$ have about twice the fraction of low metallicity stars compared to galaxies with a stellar mass of $10^{10}\,\mathrm{M_\odot}$ or $10^{11}\,\mathrm{M_\odot}$. Note that the high metallicity cutoff is essentially identical, independently of the stellar mass of the galaxies. This limit is directly connected to the metal return for high metallicity stars in TNG50. The metal fraction of the returned mass peaks at more than $30\%$ for stars with $40\mathrm{M_\odot}$ at or above solar metallicity. Therefore for a short time star particles inject gas with a metallicity of larger than $20\,\mathrm{Z_\odot}$, which sets the maximum metallicity that can be reached in the simulation.

\begin{figure*}
\includegraphics[width=0.97\textwidth]{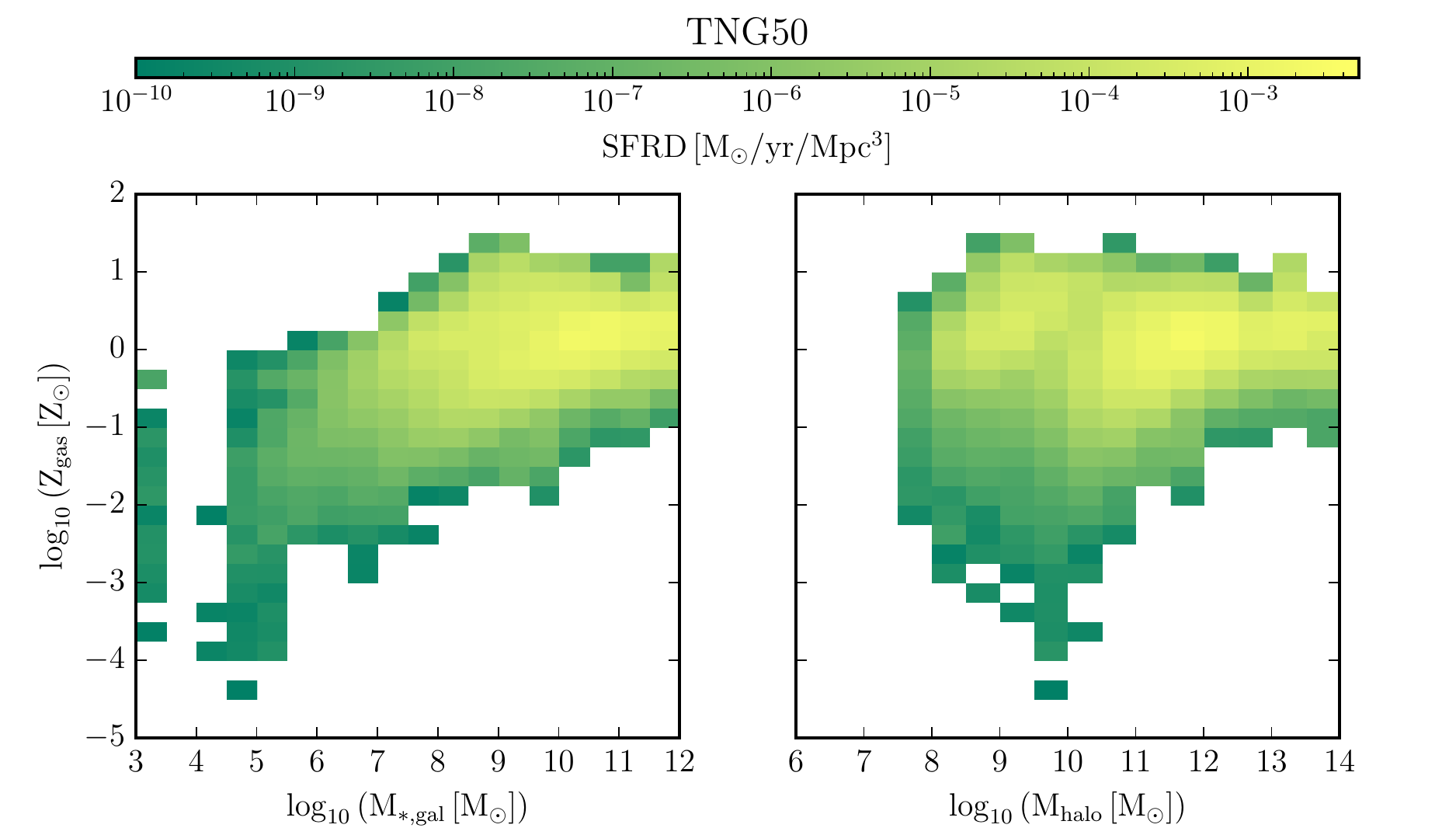}
\includegraphics[width=0.97\linewidth]{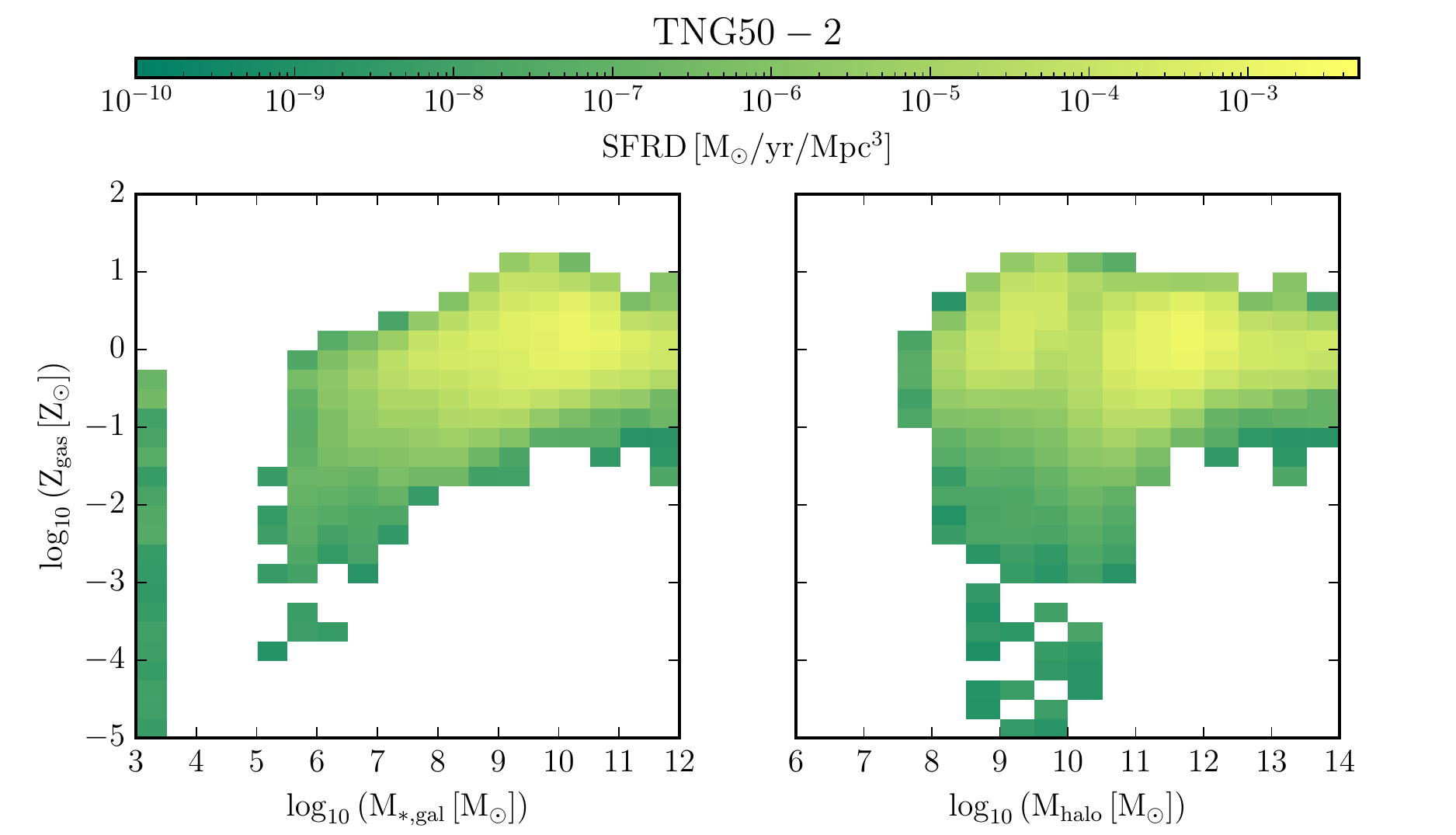}
\caption{Cosmic star formation rate density at $z=0$ binned by metallicity of  the star-forming gas and stellar mass of the host galaxy (left panel) and halo mass of the host (right panel). The top row shows TNG50, the bottom row gives TNG50-2, which has $8$ times worse mass resolution than TNG50. Galaxies without stars are set to a stellar mass of $10^3\,\mathrm{M_\odot}$. There is more starformation at extremely low metallicities in TNG50-2 than in TNG50, which emphasizes the need for high resolution simulations when looking at low metallicity star formation.}
\label{fig:SFRZeroMetallicity}
\end{figure*}

A different way of looking at the population of low metallicity stars at the present day is by considering their contribution to the total stellar mass of galaxies of a given stellar mass. We show this for stars below a metallicity of $0.1\,\mathrm{Z_\odot}$ and $0.01\,\mathrm{Z_\odot}$, respectively, in Fig.~\ref{fig:StellarMassInGalaxiesSum}. Despite hosting most of the low-metallicity stars, massive galaxies at the same time have the smallest fraction of them among their stellar populations. The increase for stellar masses larger than $10^{11}\,\mathrm{M_\odot}$ is likely an imprint of the increasing contribution of accreted stars from lower mass galaxies to the total stellar mass for these galaxies. The fraction of metal poor stars in a galaxy increases with decreasing stellar mass. For a metallicity of $Z<0.1\,\mathrm{Z_\odot}$, it increases from $1\%$ for galaxies with a stellar mass of $10^{11}\,\mathrm{M_\odot}$ to essentially $100\%$ for galaxies with a stellar mass smaller than $10^{6}\,\mathrm{M_\odot}$.

To summarise, even though low metallicity stars constitute only a tiny fraction of the stars in massive galaxies, the dominant contribution of stellar mass by massive galaxies to the total stellar mass in TNG50 means that old low metallicity stars are still more likely to be found in massive galaxies than in low mass galaxies with low average metallicities.

\section{Where in their host galaxies do low metallicity stars form and where are the old low-metallicity stars?}

\label{sec:location}

Neither star formation nor metals are distributed uniformly in galaxies. It is well known that star formation correlates with gas surface density \citep{Schmidt, Kennicutt}, which is typically higher close to the center of a galaxy. Metallicity also typically changes with galactic radius. In TNG50, the metallicity gradients are typically negative, so the lowest metallicity gas is found in the outskirts of galaxies \citep{Hemler2021}.

Thus, looking at the location of low-metallicity star formation and stars in galaxies not only gives us more information about where to find transients from low-metallicity progenitors in resolved galaxies, but also tells us about galaxy physics. In particular about the large-scale inflows and outflows of gas and mixing of metals in galaxies. Since the stellar mass of any nearby galaxy is dominated by old stars, and the formation rate of low metallicity stars declines over time as we have demonstrated, we need to look at the location of low-metallicity star formation and the typical location of low-metallicity stars separately.

We first show radial probability distributions of total and low metallicity star formation for galaxies divided into three different stellar mass ranges in Fig.~\ref{fig:profilesSF} at $z=2$ and $z=0$. In addition we show the total star formation rate of galaxies that have at least some star-forming gas with a metallicity $Z_\odot < 0.1\mathrm{Z_\odot}$.

We immediately see that most of all newly formed stars at both redshifts form at radii smaller than the stellar half-mass radius except for the most massive galaxies at $z=0$. The shift to larger radii between $z=2$ and $z=0$ is a sign of the inside-out growth of galaxies in TNG50. Typically the distribution peaks around $0.5\,\mathrm{R_{halfmass,*}}$. In contrast, low metallicity stars with $Z_\odot < 0.1\mathrm{Z_\odot}$ typically form at larger radii. For the lowest stellar mass bin at $z=2$ they peak at two to three times larger radii than the total star formation. For the intermediate stellar mass bin, low metallicity star formation reaches a plateau at $\gtrsim 1.5\,\mathrm{R_{halfmass,*}}$. The probability distribution of very low metallicity star formation continues to rise for larger radii. Note that even very low metallicity stars with $Z_\odot < 0.01\mathrm{Z_\odot}$ are also still formed at radii smaller than $0.5\,\mathrm{R_{halfmass,*}}$. These distribution are consistent with expectations from the global mass metallicity relation for galaxies in TNG and its scatter \citep{Torrey2019} and their internal radial metallicity gradients \citep{Hemler2021}.

The most massive galaxies at $z=2$ do not have any star-forming gas with $Z_\odot < 0.01\,\mathrm{Z_\odot}$ out to $R=2\,\mathrm{R_{halfmass,*}}$. They do, however, still contain star-forming gas with $Z_\mathrm{gas} < 0.1\,\mathrm{Z_\odot}$ at $r>1.5\,\mathrm{R_{halfmass,*}}$. The variance in the probability distribution here indicates that the formation of low metallicity stars at small radii is an effect of stochastic accretion of low metallicity gas into the star-forming phase.

At $z=0$ the lowest mass galaxies have a flat distribution of low metallicity star formation. Very low metallicity star formation still happens at small radii, but its distribution is patchy, again indicating stochastic accretion of very low metallicity gas into the star-forming gas phase. The flat and smooth distribution of low metallicity star formation seems intriguing and may also point to stochastic accretion being relevant.

Intermediate mass galaxies at $z=0$ have little but non-zero low metallicity star formation in the center and a probability distribution of low metallicity star formation that increases with distance from the center. This distribution again seems consistent with the expectation from global scaling laws. The distribution of low metallicity star formation for the most massive galaxies resembles the distribution of very low metallicity star formation for low-mass galaxies, with the exception that low metallicity star formation for the galaxies with $M_*>\mathrm{10^{10}M_\odot}$ only happens at radii $r>\mathrm{R_{halfmass,*}}$. The probability distribution again points to stochastic accretion of low metallicity gas driving this low metallicity star formation.

Time integration of the radial distributions modified by dynamical processes leads to the radial distribution of stellar mass $z=0$, as shown in Fig.~\ref{fig:profilesMass}. We only show the distribution for galaxies in our intermediate stellar mass bin, and only at $z=0$, because the distributions are very similar for the other galaxy mass bins and at $z=2$. The similarity between the $z=2$ and $z=0$ profiles can naturally be explained because most stars, in particular the low metallicity stars, form before $z=2$ already, so neither the stellar halfmass radius, nor the distribution of low metallicity stars changes dramatically afterwards.

The most interesting feature is that the radial distribution of stellar mass including all low metallicity stars with $Z_* < 0.1\,\mathrm{Z_\odot}$ and the distribution only including very low metallicity stars with $Z_* < 0.01\,\mathrm{Z_\odot}$ are essentially identical. This may look somewhat unexpected given the radial distribution of star formation shown in Fig.~\ref{fig:profilesSF}, but it is more easily understood when one takes into account that the total mass of low metallicity stars is dominated by stars that formed before $z=2$, as discussed before. The difference between low metallicity stars and the total population looks consistent with observed local dwarfs \citep{Genina2019}, but a more detailed comparison with many individual galaxies may be wanted.

\section{Can there be zero metallicity star formation today?}

\label{sec:zerometallicitysfr}

Finally, we turn to the most extreme low metallicity stars, i.e. the first generation of stars that are born from primordial gas that has not been enriched by any metals. The formation and fate of those Pop~III stars is still poorly understood, in part because they still have not been observed directly. Naively, they are assumed to form only at very high redshift when the first galaxies form their first stars. Those quickly enrich their host galaxies with metals so that later generations of stars already contain metals produced in stars that formed before them.

However, isolated halos could in principle assemble sufficiently late to form their first stars from primordial gas at low redshift. To investigate star formation at the lowest metallicities at $z=0$, we show the star formation rate density split by gas metallicity, and the stellar mass of the galaxy and total mass of its halo in Fig.~\ref{fig:SFRZeroMetallicity}, for TNG50 and TNG50-2. The latter simulates the same box as TNG50 but at a mass resolution that is factor of $8$ lower.

In TNG50, there are seven galaxies at $z=0$ that contain primordial starforming gas. None of these galaxies have formed any stars yet, so they are about to form their first stars at $z=0$. Their halo mass at $z=0$ ranges from $5\times 10^9\,\mathrm{M_\odot}$ to $10^{10}\,\mathrm{M_\odot}$, so they are just large enough for atomic cooling to overcome the UV background in their centers.

To understand if the history of those halos is plausible or a numerical artifact, it is important to look at the properties of their progenitor halos at $z=6$. At this time, just before their gas is heated again by reionisation, the halos had the best conditions in the past to form stars. Nevertheless, they may have been inhibited from doing so by a lack of resolution or by approximations of our galaxy formation model.

Tracing the halos back with a merger tree, we find that two of the seven halos do not yet exist at $z=6$ in the simulation, and the other five halos have halo masses of $4\times10^7\,\mathrm{M_\odot}$ to $10^8\,\mathrm{M_\odot}$, just above the minimum mass to be detectable as halos at the resolution of TNG50. In dedicated high resolution zoom simulations that include molecular cooling, the halo mass threshold to form stars before $z=6$ seems to be around a mass of $10^7\,\mathrm{M_\odot}$ at $z=6$ \citep{Simpson2013}. We conclude that it is possible that at least the two halos that do not have a progenitor in TNG50 at $z=6$ indeed assemble late enough to form stars for the first time at $z=0$.

The importance of resolution to avoid that early star formation is artificially suppressed in low mass galaxies can be demonstrated by repeating the same analysis for TNG50-2, which simulates the same box as TNG50, but at a mass resolution that is $8$ times worse. As can be seen in the bottom row of Fig.~\ref{fig:SFRZeroMetallicity}, there are many more metal-free star-forming galaxies in TNG50-2 at $z=0$ than in TNG50. More precisely, TNG50-2 has $39$ galaxies forming stars from pristine gas with a total star-formation rate of $1.6\times 10^{-2}\,\mathrm{M_\odot\,yr^{-1}}$ compared to TNG50 with $9$ such galaxies with a total star-formation rate of $7.0\times 10^{-4}\,\mathrm{M_\odot\,yr^{-1}}$.

This difference is a direct consequence of the lack of resolution at high redshift, which prevents galaxies with halo masses as high as $10^9\,\mathrm{M_\odot}$ from forming stars before $z=6$. We therefore caution to carry out such an analysis at a resolution worse than TNG50, but also emphasize again that the few metal-free star-forming galaxies we find in TNG50 may still be impacted by insufficient numerical resolution or limitations of our ISM model. Note also that the peak of the metallicity distribution of star formation at $z=0$ in massive galaxies shifts to slightly lower metallicity for TNG50-2 compared to TNG50.

In addition to not having yet formed stars themselves, the galaxies need to be isolated enough to avoid becoming polluted by more massive galaxies nearby. This is reflected in the cosmological environment of the galaxies. The background density around the seven metal-free star-forming galaxies in TNG50 on a scale of $1\,\mathrm{Mpc}$ is close to the mean density of the universe. These are some of the lowest background densities for star-forming galaxies we find in TNG50.

In this environment around the mean density the seven galaxies are able to still grow sufficiently large to eventually be able to form stars. At the same time, they are far enough away from more massive galaxies to avoid being polluted by their metal-enriched outflows. We conclude that it would be very interesting to further investigate this question with high resolution zoom-in simulations that include molecular cooling. Moreover, if Pop~III stars lead to unique transients, it may be possible that they could be found with upcoming transient surveys much closer than previously expected.

\section{Discussion}

\label{sec:discussion}

Cosmological simulations are crucial to inform us about parts of the universe that are not easily accessible to observations, e.g.~the early enrichment of galaxies at high redshift and the low metallicity tail of ongoing star formation. However, to properly understand and assess the results presented above it is important to discuss the uncertainties involved. This includes both uncertainties of cosmological simulations in general and of the  specific simulation model and numerical implementation of the IllustrisTNG project.

Here we are interested primarily in the creation and distribution of metals. The production of metals depends in our modelling on star formation and stellar evolution. They determine how many stars form, and when and how many metals are released from a stellar population of a certain mass and metallicity. Both carry significant fundamental uncertainties owing to our limited understanding of star formation and stellar evolution, and are implemented as effective models in IllustrisTNG. Importantly, the star formation model is calibrated to reproduce the observed Kennicut-Schmidt relation \citep{Springel2003}. It uses updated stellar yields \citep{TNGMethodPillepich} compared to the original Illustris simulation \citep{Vogelsberger2013}, though they still carry signifant uncertainties. Note again that the IllustrisTNG simulations are consistent with current observations constraining the global evolution of metallicity in galaxies, as seen in the mass-metallicity relation \citep{Torrey2019} for TNG100. This qualitatively validates our basic assumptions on the metal cycle in the simulations. However, one should keep in mind that significant quantitative uncertainties remain about the observed mass-metallicity relation.

The most important remaining uncertainties concern the infancy of galaxies at high redshift, were there are no direct observational constraints, so for this epoch our models are essentially extrapolations from the present-day universe. Since a large fraction of metal poor stars are formed in this phase this uncertainty is directly relevant for our results.

In addition to the creation of metals, the distribution of metals from star particles into their surrounding gas and the mixing of metals in the gas phase carry uncertainties. In TNG50, metals are distributed into the closest gas mass equal to $5\times10^{6}\,\mathrm{M_\odot}$ surrounding a star particle. This likely overestimates the amount of ISM material with which newly ejected metals are  mixed on very short timescales. Injecting metals into only a single cell increases the scatter for very rare (e.g. r-process) elements, but the effect is small for the bulk of the gas \citep{VandeVoort2020}. After the initial injection, the mixing of metals is governed by hydrodynamics, subject to some numerical diffusion due to limited spatial resolution. 

Finally, in the IllustrisTNG model we do not explicitly model local supernova feedback. Instead we employ an effective model for the pressurised ISM \citep{Springel2003}. This likely leads to an underestimation of the local turbulent velocity field and therefore possibly also to an underestimate of local turbulent mixing. In contrast, gas flows on galaxy scales are likely modelled reasonably well, since IllustrisTNG reproduces global and structural properties of galaxies. 

Finally, the uniform UV background in the IllustrisTNG model turns on instantaneously at $z=6$, rather than gradually reaching full strength \citep{TNGPublicRelease}. This may allow small galaxies to produce more stars than expected before their gas is heated at $z=6$ and they stop forming any further stars. This is particularly relevant for the question of forming metal-free stars at low redshift discussed in Sec.~\ref{sec:zerometallicitysfr}. Also, note that dust is not included in the IllustrisTNG model, and may represent a significant sink of metals \citep{Draine2007}.

With these uncertainties in mind it is interesting to take a step back and look at our results as part of a larger picture. In particular we want to stress the potential of stellar physics and galaxy formation physics constraining each other via low metallicity stars. The most recent connection has been opened up with the detection of stellar mass BH mergers with LIGO \citep{Abbott+2016_astrophysical_implications}. Merger rates of such BH binaries depend strongly on the formation history of low metallicity stars. Currently this allows us to use cosmological simulations to make predictions for merger rates of stellar mass BHs and their redshift evolution \citep[see, e.g.][]{VanSon2021,Bavera2021}. However, with the number of observed mergers of stellar mass BHs quickly increasing, we may in the near future be able to turn this connection around and use the observed merger rates of stellar mass BHs and their redshift evolution to constrain the number and formation history of low metallicity stars. This will directly constrain galaxy formation models.

Another connection that has existed for a much longer time is stellar transients that originate primarily or even exclusively from low metallicity stars. For those transients cosmological simulations can make predictions where and when to look for them \citep{Briel2021}. With the massive increase of observational data on transients expected from  forthcoming surveys like LSST, combined with a sufficient understanding of the progenitor systems of those transients, we may be able to put direct constraints on galaxy formation models.

\section{Summary and conclusions}

\label{sec:summary}

Here we have used the high-resolution cosmological simulation TNG-50 to constrain when and where metal-poor stars form. Our main results can be summarised as follows.

In Sec.~\ref{sec:global} we showed that $20\%$ of the low metallicity stars with $\mathrm{Z_{gas}}<0.1\,\mathrm{Z_\odot}$ in TNG50 still formed after $z=2$, but only $3\%$ of the stars with very low metallicity $\mathrm{Z_{gas}}<0.01\,\mathrm{Z_\odot}$. Nevertheless, star formation even at the lowest metallicities still persists all the way to $z=0$. Moreover, we find that cosmological simulations in general are consistent with at least some of the observation-based models \citep{Chruslinska2019,Chruslinksa2021}.

In Sec.~\ref{sec:hostgalaxies}, we showed that low metallicity stars with $Z_* < 0.1\,\mathrm{Z_\odot}$ are still formed at $z=0$ in TNG50 even in massive galaxies. We find that the mass of stars formed below a metallicity of $0.1\,\mathrm{Z_\odot}$ is distributed almost uniformly over host galaxies of different stellar mass. The formation of stars with metallicity $Z_* < 0.01\,\mathrm{Z_\odot}$ is limited to low-mass galaxies for $z<0.5$, but still present in massive galaxies at $z=1$. Intriguingly, most old low metallicity stars are found in massive galaxies that have an average stellar metallicty close to solar. The contribution of low metallicity stars to the total stellar mass of their massive host galaxies is negligible.

In Sec.~\ref{sec:location} we confirmed the naive expectation that low metallicity stars are generally formed at larger radii than all stars, and that their number increases with galactic radius. We find an interesting exception for low mass galaxies with $M_\mathrm{*,gal} < 10^8\,\mathrm{M_\odot}$ at $z=0$ that have a flat probability distribution for the birth radius of low metallicity stars. We also find an irregular distribution for massive galaxies with $M_\mathrm{*,gal} > 10^{10}\,\mathrm{M_\odot}$. We argued that these radial distributions of low metallicity star formation are not only useful to inform us about likely locations of transients from low metallicity progenitor stars, but may also tell us more about galaxy formation, in particular how low metallicity gas is accreted into the star-forming phase, and how long it survives before it is mixed with higher metallicity gas.

Finally, in Sec.~\ref{sec:zerometallicitysfr} we analysed seven galaxies in TNG50 that contain star-forming gas with primordial metallicity at $z=0$. We find that these galaxies live at small cosmological overdensities and are sufficiently isolated not to be polluted by larger galaxies. They just grew large enough to start cooling efficiently and form stars at $z=0$. We conclude that at least some of them seem to be physically plausible and could present an interesting target to search for in observations. Even though the total star formation rate of these galaxies is very low, if they form PopIII stars that die as bright unique transients we might be lucky enough to find them.

\section*{Data availability}
The IlustrisTNG simulations are fully publicly available at \url{https://www.tng-project.org/}. The analysis data underlying this article will be shared on reasonable request to the corresponding author. 

\section*{Acknowledgements}
We thank the anonymous referee whose comments helped to improve the paper. LvS and SdM acknowledge partial financial support from the  National Science Foundation under Grant No. (NSF grant number 2009131), the Netherlands Organisation for Scientific Research (NWO) as part of the Vidi research program BinWaves with project number 639.042.728 and the European Union’s Horizon 2020 research and innovation program from the European Research Council (ERC, Grant agreement No. 715063).
%%%%%%%%%%%%%%%%%%%%%%%%%%%%%%%%%%%%%%%%%%%%%%%%%%



\bibliographystyle{mnras}

%%%%%%%%%%%%%%%%%%%%%%%%%%%%%%%%%%%%%%%%%%%%%%%%%%

% Don't change these lines
\bsp	% typesetting comment
\label{lastpage}
\end{document}